# Advanced in silico characterization of nanomaterials for nanoparticle toxicology

*Ian Rouse, David Power, Erik G. Brandt, Matthew Schneemilch, Konstantinos Kotsis, Nick Quirke, Alexander P. Lyubartsev, Vladimir Lobaskin\**

Dr. I. Rouse, D. Power, Dr. K. Kotsis, Prof. V. Lobaskin
School of Physics, University College Dublin, Belfield, Dublin 4, Ireland
E-mail: vladimir.lobaskin@ucd.ie

Dr. M. Schneemilch, Prof. N. Quirke
Department of Chemistry, Imperial College 301G Molecular Sciences Research Hub White City Campus, 80 Wood Lane, London W12 OBZ, United Kingdom

Dr. E. G. Brandt, Prof. A. P. Lyubartsev
Department of Materials and Environmental Chemistry, Stockholm University
S-10691 Stockholm, Sweden.

**Abstract**
Nanomaterials possess a wide range of potential applications due to their novel properties compared to bulk matter, but these same properties may represent an unknown risk to health. Experimental safety testing cannot keep pace with the rate at which new nanoparticles are developed and, being lengthy and expensive, often hinders the development of technology. An economic alternative to in vitro and in vivo testing is offered by nanoinformatics, potentially enabling the quantitative relation of the nanomaterial properties to their crucial biological activities. Recent research efforts have demonstrated that such activities can be successfully predicted from the physicochemical characteristics of nanoparticles, especially those related to the bionano interface, by means of statistical models. In this work, as a step towards in silico prediction of toxicity of nanomaterials, an advanced computational characterization of these materials has been proposed and applied to titanium dioxide nanoparticles. The characteristics of nanoparticles and bionano interface are computed using a systematic multiscale approach relying only on information on chemistry and structure of the nanoparticles.

# 1. Introduction

Throughout the 21st century, Feynman's vision of the control of materials down to the atomistic level has begun to take shape in the form of nanotechnology: the study and use of materials of characteristic sizes on the order of 1 – 100 nanometers. At this scale, matter behaves significantly differently to bulk materials as a consequence of increased surface area, modified coordination of surface atoms, and different electronic band structures, amongst other properties. Consequently, these nanoparticles (NPs) may exhibit properties varying dramatically from the bulk materials, from absorbing specific wavelengths of light to increased catalytic activity. These effects may be tuned by manipulating the average size and shape of the NP, and as a result, nanoscale particles and fibers have found use in areas as diverse as food, medicine, cosmetics, and construction materials.

Yet these beneficial properties may also lead to unintended consequences when nanoparticles are inhaled, ingested, or otherwise taken into the body. Through a variety of mechanisms, nanoparticles may penetrate the cells and tissues and cause adverse effects ranging from inflammation to fibrosis or cancer.[1,2] *In vivo* and *in vitro* studies enable a direct measurement of the toxicity in model cells and organisms, but are costly, time-consuming, and are not immediately transferable from one type of NP to another, requiring that each size, shape and material is tested individually. In this situation, statistical models show great promise: quantitative structure-activity relationships (QSAR) relating the structure and basic properties of NPs to specified complex activities such as cell uptake, as well as read-across methods, where predictions are made by association, can take advantage of machine learning (ML) techniques to predict the potential risks.[3] While the entire pathway from the initial molecular interaction to the development of an adverse outcome may be too complex to be predictable based on the simplest NP descriptors, one can focus the effort on evaluating the likelihood of the molecular initiating events (MIE) or key events (KE) of the adverse outcome pathway – a sequence of biological events leading to the toxic effect. A common example of an MIE is given by lipid membrane disruption in the lung or at a cell surface or protein-mediated uptake of NPs into the cell, which may finally lead to cancer or fibrosis.[1] *In silico* methods could, in theory, allow for a direct simulation of the interaction between an NP and proteins or lipid membranes, but in practice obtaining results on meaningful timescales using atomistic models would take an infeasible amount of computational time, with even state-of-the-art techniques enabling only a few hundred nanoseconds of the system's evolution.[4] Even when these simulations are accelerated through the use of coarse-graining techniques, the direct modelling of the interaction of an NP and the body remains out of reach. A way around this obstacle is to split the task of prediction of NP activity into two parts: (i) a physics-based model that derives relevant complex properties of the NP from the basic information and (ii) a data-driven statistical model that would relate those properties with complex activities relevant for NP uptake and hazard.

By now, it has been shown that certain properties of an NP can be correlated with adverse outcomes, for example, nanofibers with a high aspect ratio are associated with frustrated phagocytosis and cancer.[5] Metal ion release rates from metal oxide NPs and conduction band gaps correlate with cytotoxicity.[6, 7, 8] The overlap of conduction band energy levels with the cellular redox potential in such systems determines the ability of NPs to induce oxygen radicals, oxidative stress, and inflammation.[7,8] Beside the hazard itself, the transport and exposure characteristics can also be related to the NP properties. Specifically, it has been shown that statistics of protein adsorption on NPs correlates with the cell uptake.[9,10] We, therefore, can aim at finding a set of descriptors – physical or chemical properties of an NP – that can be relevant for all the essential stages of the NP-induced pathway: NP transport and uptake, reactivity, persistence, and potential local hazard. These descriptors may range from simple physical properties such as the characteristic size of the NP to more complex properties such as its affinity towards a particular protein or biomolecular fragment in a specific biological medium.[11] By associating these descriptors (or their combinations) with MIEs or KEs of given adverse outcome pathways via QSARs, the toxicity of a novel nanomaterial may be predicted simply by evaluating its relevant characteristics and supplying them to the statistical model. Some of the properties, such as hydrophobicity, protein adsorption affinity, dissolution rates and the ability to generate reactive oxygen species, were identified by the nanosafety expert community as toxicity determinants,[12] and their provision is seen as a crucial step towards the development of predictive schemes. Where such properties are not available experimentally, materials modeling can be useful to provide the necessary data.

In this work, we present details on how a range of quantitative descriptors of NPs and the bionano interface can be calculated for an arbitrary inorganic nanomaterial from first principles using coupled quantum chemistry, atomistic molecular dynamics and mesoscopic methods. These descriptors provide immediate insight into the physical and chemical properties of the given NPs and further form the basis for machine learning and statistical techniques for toxicity prediction, e.g. as input for QSAR models covering the existing corpus toxicological data. As a demonstration of this method, we present results calculated for both the anatase and rutile forms of titania ($TiO_2$), representing the most commonly produced nanomaterials, of a range of sizes and surface charges to demonstrate the broad applicability of the method.

## 2. Results

We start with evaluation of intrinsic properties of the target nanomaterial, in this case, titania using common electronic structure methods. In what regards the extrinsic properties, we make an attempt to improve the atomistic model of the interface of the material with water and thus provide more accurate and more advanced description of the bionano interactions. To compute the more advanced characteristics of NPs we use the recently proposed multiscale approach[13,14] which enables modelling of large molecular assemblies in the length and time domain not easily reachable by atomistic simulations and allows us to relate the atomic structure and basic descriptors of the material to bio-nano interactions. Our multiscale method includes:

- parameterization of the atomistic force-field for the material using electronic structure methods
- calculation of interactions of the biomolecule building blocks (amino acids, lipid segments, DNA bases) with the surface of the material and interaction between the building blocks at the atomistic level at the specified conditions
- parameterization of the CG force field for biomaterial building blocks and construction of the sample of arbitrary size and shape
- CG modelling of interaction of entire biomolecule with the biomaterial surface and calculation of preferred orientation and the mean adsorption energy

In Section 2.1 we consider intrinsic properties of the NP, that is, those that depend only on the morphology and material of the nanoparticle. In Section 2.2 we describe the coarse-graining scheme in detail and present results for the binding of proteins to titanium dioxide nanoparticles. In Section 2.3 we present descriptors regarding the interaction of the NP with lipids to parameterize their interaction with cell membranes.

### 2.1. Structural NP descriptors and intrinsic properties

The simplest descriptors one can calculate reflect basic geometrical properties of the NP defined in terms of the primary size $R$, where $R$ corresponds to the radius of a spherical or cylindrical NP, and for cubic NPs we take the length of a side to be $2R$. For cylindrical NPs we also require the total length $L$, which we typically assume to be much larger than $R$. From these parameters, the surface area $A$ and volume $V$ may be straightforwardly calculated: $A = 4\pi R^2, 2\pi RL + 2\pi R^2, 24R^2$ and $V = 4/3\pi R^3, \pi R^2 L, 8R^3$ for the sphere, cylinder and cube respectively. An NP with a large surface area can reasonably be expected to bind to a greater number of proteins, while an NP with a large volume will exhibit a greater van der Waals (vdW) attraction to other particles and so may bind to these particles more strongly. The size and shape of the NP will also influence factors such as the speed at which it diffuses in biological media, whether it can dock into binding regions of proteins and whether it can pass through pores. Further useful descriptors can be derived from the atom configurations and energies, once the crystal structure of the NP is known.[15] They include molecular masses, coordination numbers, and energies of atoms in the bulk and at the surface, number of oxygen or metal atoms. All these quantities can reasonably be expected to determine biological activity of the NP.

Electronic properties of a specific nanomaterial can be obtained through computational techniques such as quantum mechanical semi-empirical calculations based on the Hartree–Fock formalism[16] and density functional theory methods, with results for $TiO_2$ shown in **Table 1** and Table S1 of the Supporting Information. This allows for both the optimization of the structure of a given nanomaterial on the density functional tight binding (DFTB) level[17] and the density functional theory (DFT) level and the calculation of physicochemical properties of this material by semi-empirical quantum mechanical calculations.[16] Temperature and the size of the NP are important factors that determine the stability of the $TiO_2$ forms. Here, we calculate electronic band gaps, ionization potentials, electronegativity, hardness, dispersion corrections, polarizability and the dipole moment for bulk anatase and rutile representing the core of a $TiO_2$ NP at the semi-empirical level of theory in comparison to DFT and DFTB calculations for band gaps and ionization potentials. Such descriptors have been applied in statistical models to describe toxicity of metal oxide nanomaterials in relation to the core properties of a NP.[8,15,18] Moreover, we employ DFT calculations (Table S1 in Supporting Information) using the SIESTA code[19] where the unit cells of anatase and rutile $TiO_2$ are used to describe the extended bulk structures by Monkhorst-Pack meshes for the point sampling of the Brillouin zone integration.[20]

**Table 1**: Material properties calculated through MOPAC [16], DFTB+ [17], and SIESTA[19] for the anatase and rutile forms of $TiO_2$.

| $TiO_2$ Solid systems | Band gap (eV) | Ionization potential - Valence band maximum energy (eV) | Mulliken Electro-negativity | Parr & Pople absolute hardness | Dispersion energy per atom (kJ mol$^{-1}$) | Polarizability (Å$^3$) | Dipole moment (Debye) |
|---|---|---|---|---|---|---|---|
| Semi-empirical | | | | | | | |
| Anatase | 9.41 | 6.67 | 1.96 | 4.70 | -4.92 | 133.01 | 0.34 |
| Rutile  | 9.66 | 5.51 | 0.68 | 4.83 | -5.41 | 122.32 | 6.16 |
| DFTB | | | | | | | |
| Anatase | 3.42 | 3.36 | - | 1.45 | - | - | - |
| Rutile  | 2.68 | 3.64 | - | 2.18 | - | - | - |
| DFT | | | | | | | |
| Anatase | 2.49 | 8.75 | - | - | - | - | - |
| Rutile  | 2.25 | 9.46 | - | - | - | - | - |

## 2.2. Extrinsic NP properties. Bio-nano interactions descriptors.

It is well-known that a NP immersed in a biological medium forms a corona of adsorbed proteins, lipids and sugars, and that the composition of this corona is highly dependent on the affinity of each type of protein and lipid present to the NP.[21] Thus, the energy of adsorption, also referred to as the binding energy, of biomolecules to an NP is an important set of descriptors characterizing interactions of NPs with biomolecules. To calculate these values, we employ a coarse-graining approach in which the interactions between small biomolecules, e.g. amino acids, and the NP surface are evaluated using atomistic molecular dynamics to obtain potentials of mean force (PMFs). These potentials, together with additional terms describing the electrostatic potential and long-range van der Waals attraction, are used as the input for a calculation of the binding energy of proteins built up from these smaller fragments. In this way, parameterizing a small number of building blocks is sufficient to evaluate the strength of binding between NPs and proteins for which an atomistic molecular-dynamics simulation would be prohibitively time-consuming.

The first step in this parameterization is developing a force field for atomistic simulations between biomolecular fragments and the NP. The quality of the force field used in classical atomistic simulations is of primary importance for the simulation to produce reliable results.

While there exist validated force fields describing bulk materials or aqueous solutions of organic and biomolecules, they are less developed for description of surface properties of (nano)materials. For modeling of metal or metal oxide surface in aqueous media an additional problem is adequate representation of the material surface which is modified (in comparison to the bulk material structure) by reactions with water building a surface-specific hydroxylated layer containing hydroxyl groups, bound molecular water or other surface modifications. This issue was handled as discussed in the Methods section and a summary of the identified force field types and force field parameters are given in **Table 2** and **Table 3**.

**Table 2**: Non-bonded force field parameters for $TiO_2$ in water environment

| Atom type | Coordination | Description | Partial charge (e) | σ, (Å) | ε, (kJ mol$^{-1}$) |
|---|---|---|---|---|---|
| H | H -O1 | hydrogen | 0.417 | 0 | 0 |
| TiA | Ti -O6 | Bulk Ti | 2.248 | 1.99 | 13.79 |
| TiB | Ti - O5 | Surface Ti / defect | 2.159 | 1.9 | 13.79 |
| OA | O - Ti3 | Bulk $TiO_2$ oxygen | -1.124 | 3.51 | 0.409 |
| OB | O - Ti2 | Bridge oxygen | -1.035 | 3.42 | 0.401 |
| OF | O – H1Ti1 | Hydroxyl oxygen | -0.913 | 3.29 | 0.389 |
| OG | O - H1Ti2 | Protonated oxygen bridge | -1.035 | 3.42 | 0.401 |
| OH | O - H2Ti1 | Adsorbed water | -0.923 | 3.151 | 0.634 |

**Table 3**: Bonded force field parameters for TiO$_2$ in aqueous environment

| Bond | Equilibrium distance, Å | Force constant, kJ mol$^{-1}$Å$^{-1}$ |
| --- | --- | --- |
| Ti* - OA | 1.9 | 8000 |
| Ti* - OB | 1.9 | 8000 |
| Ti* - OF | 1.9 | 8000 |
| Ti* - OG | 1.9 | 8000 |
| Ti* - OH | 1.8 | 400. |
| OF – H | 1.0 | 3267 |
| OH – H | 1.0 | 3267 |
| Angle | Equilibrium angle, deg | Force constant, kJ mol$^{-1}$deg$^{-1}$ |
| OF – Ti* - OH | 90. | 500. |
| OH – Ti* - OH | 90. | 500. |
| Ti* - OF – H | 114.85 | 543. |
| Ti* - OG – H | 112.6 | 564. |
| Ti* - OH – H | 114.85 | 543. |
| H – OH – H | 104.2 | 628. |

As observed in *ab initio* simulations,[22] undercoordinated surface Ti-atoms have either adsorbed water or hydroxyl groups bound to them. Each hydroxyl group contributes a charge of about $-0.4e$. We selected the fraction of hydroxyl groups at the surface as 30%, from the condition that the experimentally measured surface charge of TiO$_2$ NPs at neutral pH is about $-0.62e$ nm$^{-2}$.[23]

To enable the calculation of the binding energy for a range of complex biomolecules, we have selected a set of 30 small molecules which represent all typical molecular fragments present in biomolecular fluids, and calculated PMFs and binding energies for this set (see Methods section for details and the Supporting Information for plots of the PMFs). The list of chosen biomolecules includes:

- Amino acid side chain analogues
- Glycine and proline aminoacids with the backbone fragment (GLY, PRO)
- Modified charged forms of aminoacids with pKa values between 4 and 10 (HIS+, GLU-protonated, CYS-, denoted HIP, GAN, CYM)
- Segments of the most abundant lipids: choline (CHL) and phosphate (PHO) group of phosphatidylcholine (PC) lipids; amino group (ETA) of ethanolamine lipids (PE), ester group (EST)
- d-glucose representing sugars (DGL)

There are 20 naturally occurring amino acids of which the proteins of living organisms consist. Each full amino acid contains a peptide backbone fragment which is common to all amino acids. In order to avoid redundancy, we excluded the backbone fragment and considered the side chain analogues for all amino acids excluding glycine (for which the side chain analogue is just an H atom) and proline which has a different structure. This set of side chain analogues consists of 18 molecules. Histidine exists in two isomeric forms (denoted as HIE and HID) and we include both of them. These side chain analogues also have the same structure as hydrophobic lipid tails and certain common lipid head groups (phosphatidylserine, PS and sphingomyelin, SM), further extending the range of larger biomolecules covered by this set. The list of molecules introduced here covers all the main types of chemical entities: hydrophobic, polar, aromatic, and charged, and represent all typical molecular fragments present in biofluids. The set of binding free energies of these molecules makes a "fingerprint" of NP with respect to bio-nano interactions and so such a set of descriptors is essential in the predictive scheme of toxicity assessment.

As an example of such a bio-nano fingerprint, we computed binding free energy of these molecules to rutile (110) and anatase (101) plane surfaces. These surfaces are the lowest energy surfaces in the respective forms of $TiO_2$, and by this reason one can expect that they represent most of the surface of the NPs. A similar model has been used previously to study adsorption of proteins on a surface of rutile NP. [24] We extend the previous calculations by covering a wider range of biomolecular fragments and using a more advanced force field accounting for the difference between bulk and surface titania, allowing multiple chemical environments for oxygen atoms, and taking into account structural differences between the anatase and rutile forms. The computed binding free energies are shown in **Figure 1** and tabulated in the supplementary materials. One can clearly see differences in the adsorption profile for the two forms of $TiO_2$. The most strongly binding molecules for anatase are glutamic acid (GLU), cysteine anion (CYM) and aspartic acid (ASP), all these are negatively charged molecules with either carboxyl group or thiolate. None of these molecules show significant binding to rutile. The difference in binding could originate in the specific structure of the anatase surface where negatively charged groups of the molecules can coordinate favorably with hydrated positively charged titanium atoms. On the rutile surface, access to titanium atoms is screened by the bridging oxygen atoms, thus preventing the binding of anionic molecules. The different character of binding of small molecules have consequences for the binding of larger biomolecules and formation of NPs corona.

**Figure 1**: Adsorption free energies of biomolecule fragments to rutile (left) and anatase (right) TiO$_2$ surfaces.

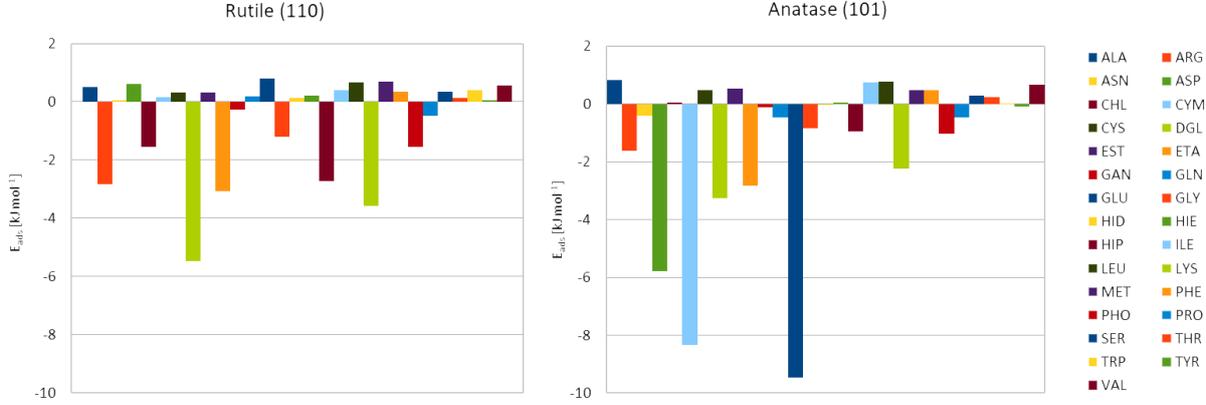

To demonstrate this, we also calculate the binding energy for larger biomolecules assembled from these fragments. The PMFs include only interactions between the biomolecule and the nanomaterial up to a distance of 1.2 nm. The bulk of the nanomaterial interacts with these molecular fragments via the long-range van der Waals interaction which may remain significant at distances beyond the cut-off used for the calculation of the PMFs. To include these effects while avoiding the need to explicitly sum over the interaction between each atom in the NP and in the target biomolecule, we employ the continuum approximation and use the Hamaker approach to write the bulk interaction between two spheres of radius $R_1$, $R_2$ with surface-surface distance $d$ as, [25,26]

$$U(d) = \frac{-A_{21}}{6}\left(\frac{2R_1R_2}{(2R_1+2R_2+d)d} + \frac{2R_1R_2}{(2R_1+d)(2R_2+d)} + \ln\ln\frac{(2R_1+2R_2+d)d}{(2R_1+d)(2R_2+d)}\right), (1)$$

where the quantity $A_{12} = \pi^2\lambda\rho_1\rho_2$ is referred to as the Hamaker constant and is a measure of the magnitude of the long-range dispersion interaction based on the number density of the two materials $\rho_1$, $\rho_2$ and the vdW interaction strength $\lambda$. A more rigorous derivation of the dispersion interaction including the effects of the medium in which these particles are immersed is achieved through Lifschitz theory.[26] For materials denoted $i = 1,2$ interacting through a medium $i = 3$, the constant $A_{12}$ must be replaced by another one $A_{123}$ expressed in terms of the refractive indices (at visible wavelengths) of the materials $n_i$, their dielectric constants $\varepsilon_i$ and the main electronic absorption frequency $v_e$ (in the UV) for material 2: [26]

$$A_{123} = \frac{3}{4}k_B T \frac{(\varepsilon_1-\varepsilon_3)}{(\varepsilon_1+\varepsilon_3)}\frac{(\varepsilon_2-\varepsilon_3)}{(\varepsilon_2+\varepsilon_3)} + \frac{3h\nu_e}{8\sqrt{2}} \frac{(n_1^2-n_3^2)(n_2^2-n_3^2)}{(n_1^2+n_3^2)^{\frac{1}{2}}(n_2^2+n_3^2)^{\frac{1}{2}}\left((n_1^2+n_3^2)^{\frac{1}{2}}+(n_2^2+n_3^2)^{\frac{1}{2}}\right)} \quad (2)$$

Clearly, this value will be different for each biomolecule that may interact with the NP. The calculated values for anatase and rutile nanomaterials with twenty common amino acids with are detailed in **Table 4** using parameters found from the literature. [27,28]

**Table 4**: Hamaker constants (in units of kJ·mol$^{-1}$) describing the bulk interaction between anatase and rutile with twenty common amino acids.

| AA | ALA | ARG | ASN | ASP | CYS | GLN | GLU | GLY | HIS | ILE |
|---|---|---|---|---|---|---|---|---|---|---|
| A(anatase-AA) | 19.53 | 23.47 | 25.28 | 25.88 | 24.88 | 23.87 | 22.86 | 24.88 | 25.89 | 16.86 |
| A(rutile-AA) | 17.19 | 21.90 | 22.30 | 22.83 | 21.94 | 21.05 | 20.15 | 21.94 | 22.83 | 14.87 |
| AA | LEU | LYS | MET | PHE | PRO | SER | THR | TRP | TYR | VAL |
| A(anatase-AA) | 16.65 | 20.13 | 22.25 | 24.68 | 18.82 | 24.27 | 20.33 | 29.42 | 22.04 | 17.07 |
| A(rutile-AA) | 14.66 | 17.74 | 19.61 | 21.77 | 16.57 | 21.41 | 17.92 | 25.97 | 19.43 | 15.03 |

Once the PMFs and Hamaker constants for the set of amino acids and other fragments of interests have been calculated, these can be used as input to calculate the binding energy for biomolecules composed of these fragments.[14] Here, we use the UnitedAtom package[14] to calculate the binding energies of a set of reference proteins (see Supporting Information) onto spherical titania NPs. In this model, the protein is represented as a set of beads, with each bead representing one amino acid. The interaction potential between the NP and a bead consists of three components. The first is a potential of mean force (PMF) describing the short-range potential obtained through atomistic simulations as described in the previous section and corrected to take into account the radius of the NP by applying a distance-dependent scaling factor. [14] Using this correction, a set of PMFs calculated for a planar slab of the material may be applied to all spherical NPs of this material, substantially reducing the computational time required to evaluate the binding energy for a set of NPs of the same material. To account for the bulk of the NP beyond the cutoff range of the PMF, the van der Waals interaction is added as the second component. This term is corrected to exclude the volume of the NP sufficiently close to the AA that it would be included in the PMF. The final component is the electrostatic interaction in the Debye-Hückel approximation, which accounts for the interaction between the surface charge of the NP and charged residues and is specified in terms of the surface (zeta) potential. The total potential for a given bead type is calculated by summing over these three contributions, and then summed over all beads to produce the total interaction potential at a given orientation of the biomolecule relative to the surface of the NP, denoted $U(z,\phi,\theta)$. The binding energy for a protein on a spherical NP of radius $R$ as a function of the orientation of the protein $\phi,\theta$ is given by, [13]

$$E(\phi, \theta) = -k_B T \ln \left[ \frac{3}{(R+a)^3 - R^3} \int_R^{R+a} z^2 \, exp\left(\frac{-U(z,\phi,\theta)}{k_B T}\right) \, dz \right], (3)$$

where $a$ is a function of $\phi, \theta$ and gives the maximum distance between the surface of the NP and the centre of mass of the protein, beyond which the protein is deemed to be unbound. Performing a Boltzmann-weighted average over orientations produces the mean binding energy,[13]

$$E_{ad} = \frac{\int\int P(E,\phi,\theta)E(\phi,\theta)d\phi d\theta}{\int\int P(E,\phi,\theta)d\phi d\theta} \quad (4)$$

with the weighting function $P(E,\phi,\theta) = \sin\sin\theta \; exp\, exp\, (-E(\phi,\theta)/k_B T)$. The resulting binding energies for a selection of the proteins considered are presented in **Figure 2** for anatase (a) and rutile (b) NPs of radius 50 nm, and are calculated under the assumption the NP surface can be adequately described by the PMFs calculated for the 101 surface for anatase and 110 surface for rutile. These values depend strongly on the radius of the NP and on a lesser extent on the value of zeta potential as depicted in Figure 2 (c) and (d). The full set of binding energies for the 96 proteins calculated on 10 radii and 5 zeta potentials is available as Supporting Information.

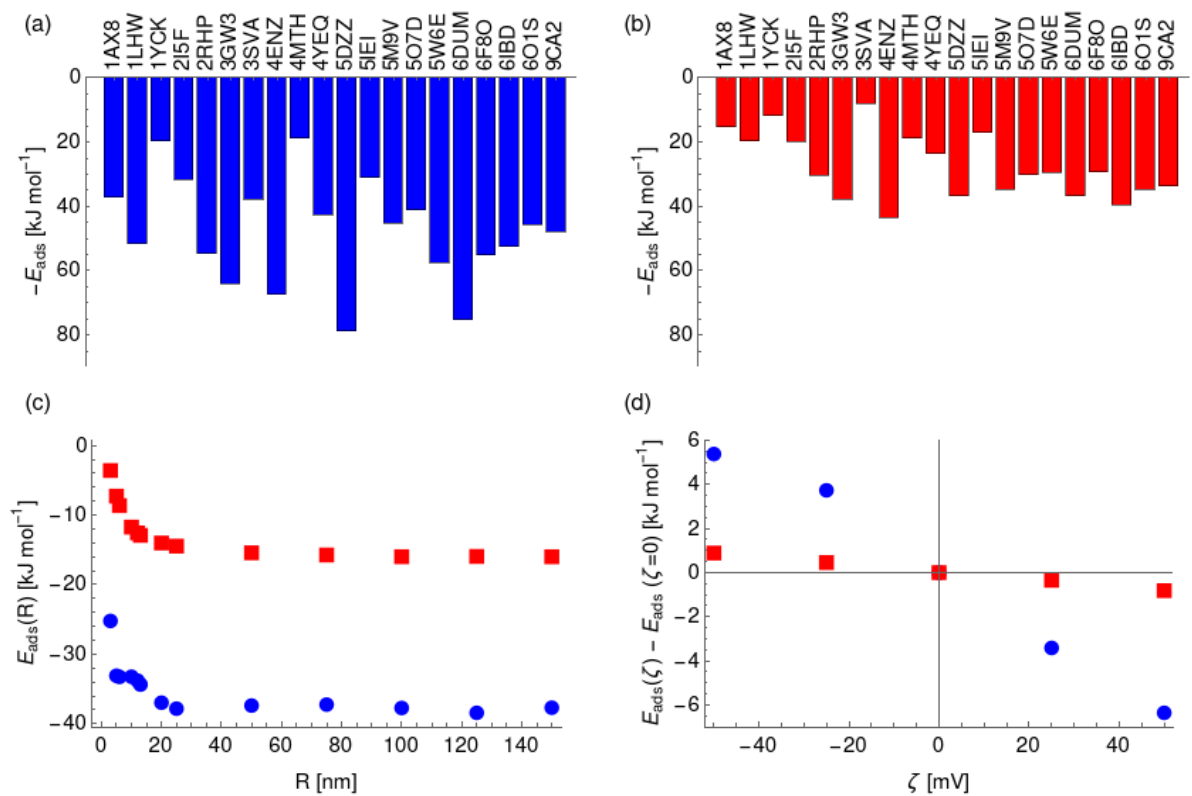

**Figure 2**: Adsorption free energies of a set of proteins (labeled by their Protein Databank identifier) onto spherical anatase (a) and rutile (b) NPs of radius 50 nm and zero zeta potential. The variation as a function of the radius for the protein with PDB identifier 1AX8 is shown in (c) and variation as a function of zeta potential for a particle of radius 50 nm in (d) for both anatase (blue) and rutile (red).

### 2.3. Immersion and adhesion enthalpies, lipid wrapping

A key measure of the degree of interaction between NPs and membranes is the extent to which the NP adheres to or is wrapped by the membrane. According to the Helfrich membrane model[29] applied by Deserno and Gelbart[30] to the wrapping of spherical particles by membranes, the outcome is determined by whether the adhesion strength $w$ (the free energy of adsorption) is sufficient to overcome the bending energy penalty associated with the required membrane deformation. Particles smaller than certain critical size will not adhere to the membrane. Larger particles will adhere and undergo wrapping; the extent to which wrapping occurs is determined by the membrane tension. With low tension and/or strong adhesion, particles will be completely engulfed, following which the particle may detach from the membrane leaving a membrane pore through which cytosol leakage can occur. The adhesion strength drives the particle wrapping process, so that any assessment of the potential nanotoxicity of a material must include an estimate of its adhesion strength $w$ for a given membrane, which may be approximated by the adhesion strength with respect to the membrane lipids. Adhesion strength data can then be used to estimate the probability of passive NP uptake by human cells as well as pulmonary surfactant disruption due to the interaction with NPs in the alveolar spaces of lungs. A major obstacle to progress is the lack of quantitative experimental data for NP/lipid bilayer adsorption. We have previously proposed methods of calculating *w* from molecular simulation and applied these to a range of materials including gold,[31] silica[32] and titania.[33] Some data for DMPC lipids on titania surfaces are shown in **Table 5.** These are calculated from atomistic molecular dynamics methods using the new force field, and so enable a comparison to the fragment-based model.

**Table 5:** A summary of the calculated heats of immersion, adsorption energy for single DMPC molecules and adhesion strength for DMPC bilayers on a range of titania surfaces.

| Polymorph | Rutile | Rutile | Rutile | Rutile | Anatase | Anatase | Anatase | Anatase |
|---|---|---|---|---|---|---|---|---|
| Miller index | 110 | 100 | 101 | 001 | 101 | 100 | 001 | 110 |
| Heat of immersion (mN m$^{-1}$) | 1024 | 1009 | 847 | 1102 | 774 | 1176 | 392 | 1006 |
| Single lipid adsorption energy (kJ mol$^{-1}$) | -1.8 | -3.4 | -13.8 | -0.4 | -1.7 | -0.2 | -0.1 | -1.9 |
| Adhesion strength (mN m$^{-1}$) | -1.8 | -3.6 | -4.0 | -0.3 | -1.0 | -1.1 | -0.6 | -4.0 |

## 3. Discussion

*In silico* material characterization provides valuable information on the NP properties that may not be readily available from experiments. Yet, due to principal technical limitations such as the large required system size or time scale, the computations at the nanoscale are necessarily approximate and often cannot guarantee quantitative accuracy in reproducing experimentally observed properties in absolute terms. The possible mismatches with experimental data may be caused by unrealistic assumptions about the material such as its crystalline order, structure of the NP surface, material purity, or the coating. Unfortunately, such assumptions are unavoidable as experimental datasheets are often lacking the relevant information.

In our calculations of intrinsic properties of titanium oxides, we used common quantum mechanical methods on anatase and rutile of bulk TiO$_2$. The methods show different accuracy with respect to different material properties. It is known that the band gap of rutile is lower than anatase TiO$_2$ by ca. 0.2 eV [34] and the ionization potential is larger for anatase than rutile. Our DFT and DFTB band gap results show the trend consistent with experiments, while semi-empirical calculations predict the opposite trend. For the ionization potential, the semi-empirical calculations give the trend consistent with experiments, while both DFT and DFTB show larger values for rutile. More accurate approaches have been used in literature to reproduce not only the relative trends between rutile and anatase TiO$_2$ but accurate values of band gaps and band alignment in accordance with measurements.[34]

In the evaluation of extrinsic properties of titania NPs, such as interaction with water, lipids and proteins, we used the new force field developed in this work. All titania surfaces exhibited a large degree of hydrophilicity as reflected in the uniformly exothermic heats of immersion, which were an order of magnitude greater than the silica surfaces studied previously. A far greater range in heat of immersion was observed on the anatase surfaces than for rutile. From the data on the lowest energy surfaces, (110) for rutile and (101) for anatase, rutile appears to be more hydrophilic. In addition, we measured the free energy of adsorption of single lipids on the titania surfaces. For the majority of the surfaces, energy of adsorption was less than the thermal energy indicating that lipids do not spontaneously adsorb on these surfaces. The only surface to display strong adsorption was the rutile 101 surface with a minimum of −13.8 kJ mol$^{-1}$. The average adhesion strength across the cleavage planes for rutile and anatase is relatively weak at -2.0±0.4 mN m$^{-1}$. However, rutile has two (100 and 101) high energy surfaces (-4 mN m$^{-1}$), while anatase has only one (110). Since particle surfaces are expected to comprise a range of low energy cleavage planes, this observation suggests a slightly greater tendency for bilayers to wrap rutile NPs compared to anatase NPs, but both forms wrap less than amorphous silica. These results indicate that the adsorption energy of free lipid molecules calculated using the coarse-grained model does not necessarily predict the adhesion energy of lipid bilayers calculated using atomistic simulations, highlighting the requirement to calculate the interaction of NPs with both larger-scale structures as well as fragment-based calculations. As discussed in Section 2.2, the anatase surface is more strongly binding to the amino acid molecules than the rutile surface is, and this in turn leads to a difference in the calculated binding energies for the proteins considered as can be seen in Figure 2, providing further evidence that a wide range of biomolecules should be selected for the calculation of predictive descriptors. The stronger binding of proteins to anatase correlates with the weaker binding of water as compared to rutile, as seen from the heats of immersion. The observed trends for the protein binding energies here are in agreement with experiments showing that anatase titania binds blood serum proteins more strongly than the rutile polymorph.[35]

As discussed previously, [14] the UA model contains a number of approximations, which may cause systematic errors in the protein adsorption energies. The most significant of these is the assumption that all contributions to the NP-protein potential can be treated additively and that the orientation of AA side chains can be neglected. Moreover, charge regulation in both the protein and the NPs are neglected, and the protein is assumed to be fixed to its native structure and cannot relax due to binding to the surface of the NP. Finally, we note that the Hamaker constants required are typically not available ab initio and must be obtained from the literature, with different sources providing different values for these constants. Thus, the calculated adsorption energies do not necessarily predict the correct absolute value. We expect, however, that these factors are not significant due to the characteristic sizes of the NP-protein complex, and we expect it to produce the correct ranking of proteins by affinity to a particular NP. This, in turn, should enable the correct ranking of the corona abundances. Moreover, the simplifications in this model present substantial time savings in comparison to more computationally intense calculations of protein adsorption using atomistic simulations.[4, 36] These atomistic simulations are limited to providing binding energies for a single NP, whereas the method outline here enables the rapid calculation of binding energies for a whole class of NPs of the same material but varying sizes, shapes, and zeta potentials.

The set of descriptors given here is by no means definitive but are selected as a set of reasonably simple properties that can be evaluated using standard computational techniques and as a basis for generating further descriptors. As an example, we note that the binding energies calculated here may be of use in estimating the composition of the protein corona which forms around a NP by using these as input for simple models of the steady-state corona.[37,38] As a further extension, we intend to use the calculated binding energies and other simple physicochemical descriptors as input for a QSAR model for the prediction of toxicological properties. Here, we have focused on titanium dioxide as this is a particularly important nanomaterial, which is being extensively covered by toxicological studies,[39] however, we stress that the methodology used here is completely general and can be applied to a wide range of nanomaterials other than titanium dioxide, e.g. gold, silica and carbon-based nanomaterials, with results for these materials obtained and to be presented in future work. Likewise, the range of biomolecules that can be evaluated can readily be extended by identifying further basic structures and generating PMFs and Hamaker constants for these.

## 4. Conclusion

In this work, we have evaluated a set of descriptors – physical and chemical properties of NPs – that might be most relevant for describing the bionano interface, and thus can be used for predicting the toxicological behavior of novel nanomaterials. These range from simple properties such as the available surface area to the affinity of specific proteins to the NP calculated using a coarse-grained model. Importantly, the described methodology does not rely on any experimental parameters and connects the advanced material properties to the basic descriptors. The proposed descriptors have been calculated for a range of titanium dioxide NPs and may be applied to arbitrary NPs, paving the way for deep-learning approaches to identify both which descriptors are most important for the prediction of toxicity and the rapid assessment of new nanomaterials.

## 5. Methods

**Semi-empirical, DFT and DFTB**

Semi-empirical calculations have been performed with the MOPAC computational code.[16] All material properties (band gap, ionization potential, Mulliken electronegativity, absolute hardness, dispersion, polarizability and dipole moment) were obtained with the PM6 method using the D3 correction on the DFT optimized structures.[40] Geometry optimizations of the rutile and anatase $TiO_2$ bulk structures were performed with SCC-DFTB[41] using the DFTB+ software[17] with the parametrization tiorg-0-1[42] and with the SIESTA code[19] with the PBE functional[43] and a DZP (double-z polarized) basis. Troullier–Martins pseudopotentials[44] were applied on the core electrons.

**Force Field Parameterization**

To parameterize the forcefield for the titanium dioxide we used a multiscale approach, where the detailed structure of the hydroxylated layer of a metal oxide nanomaterial, as well as parameters of the force field describing interactions of the fully hydrated surface with surrounding biomolecular solution are obtained from high-quality *ab initio* molecular dynamics (MD) simulations. *Ab initio* MD simulations provide a representative set of snapshots correctly representing thermal fluctuations of the studied systems, which is used for the further analysis. The method is based on partitioning the quantum mechanical electron density into atomic basins. We apply the population analysis method by Manz [45] to partition the electron density among the atoms and extract individual net atomic charges and atomic volumes for individual atoms, as well as information on the bond orders for individual pairs.

First, the net atomic charge (NAC), defined by the DDEC6 partitioning method [45]

$$q = z - N = z - \int d^3 r w(r) n(r) \qquad (5)$$

Here $z$ is the atomic number and $N$ is the number of electrons assigned to the atom. $n(r)$ is the total electron density, and $w(r)$ is the spherical weight attributed to the atom by the partitioning method. NACs provide partial atom charges which are routinely used to model electrostatic interactions in molecular simulations with empirical force fields.

Second, the cubed atomic moment (CAM)

$$V = \langle r^3 \rangle = \int d^3 r r^3 w(r) n(r) \qquad (6)$$

corresponds to the volume occupied by the atom in the material and is proportional to the local polarizability.[46]

Third, the concept of bond order (BO), defined by [47]

$$D_{ij} = 2 \int \int d^3 r d^3 r' \frac{w_i(r) w_j(r')}{w(r) w(r')} n(r) n^{DXH}(r, r') \qquad (7)$$

which quantifies the amount of shared electron density between atoms $i$ and $j$. Here, $w(r) = \sum_i w_i(r)$ is the total spherical weight, $n^{DXH}(r, r')$ is a normalized probability distribution (over $r'$ such as to exclude exactly one electron) that quantifies the so-called dressed exchange hole. The electron density close to the nucleus is depleted due to the exchange interaction. This concept can be modified for bond order by contraction/expansion of the density to align bond orders with the conventional view. With this definition the bond order between two atoms decreases smoothly to zero as the distance between them approaches infinity.

As a training ensemble, we have used trajectories generated in ab-initio molecular dynamics trajectory simulations described previously.[22] Briefly, six fully hydrated TiO$_2$ surfaces (rutile 110, 100 and 001, and anatase 101, 100 and 001) were simulated with DFT computed forces during 50 ps, and 20 snapshots for each surface taken each 1 ps from the last 20 ps of the simulations were taken for the analysis. Each member of the ensemble was subject to atom-in-molecule analysis with the Density-Derived and Chemical (DDEC6) method[46] to determine atom-in-material net atomic charges, volumes, and bond order parameters using ChargeMol v3.5 software.[48]

We have used bond order to determine whether atoms are bound, by setting a threshold value 0.25. The local connectivity of atoms was used to determine the force field types. Disregarding atom coordinations which were observed in less than 1% cases, we identified two force field types for Ti-atoms, five types of oxygen atoms, and one type of hydrogen, see Table 1. For each type of atom, we have computed average net atom charges (1) and cubic atomic volumes (2). Computed net atomic charges were used directly to determine partial atom charges of the force field, with a minor modification providing total zero charge for stochiochemical sample of $TiO_2$. Net atomic volumes were used to determine parameters of the Lennard-Jones potential using the theory developed in [49], as numerically computed and tabulated in[50,51]. We have used these values here to obtain dispersion coefficients of the Lennard-Jones potential from the atomic volumes defined by Eq. (6), as well as repulsion parameter in the Lennard-Jones potential. For hydrogen, we assumed zero Lennard-Jones parameters, to make it compatible with TiP3P water model.

**Metadynamics**

Details of how the metadynamics calculations were performed were previously reported in [14] and are summarized in the Supporting Information.

**Coarse-grained molecular model**

Binding energies for proteins on NPs were calculated using the UnitedAtom tool.[14] The interaction energy for a protein molecule with a NP is calculated as a sum of three terms: amino acid – NP surface interaction, which is calculated using the PMFs obtained in full atomistic metadynamics calculations, amino acid – bulk NP interaction evaluated by integration of the van der Waals force of the remaining volume of the NP, and electrostatic interaction obtained via screened Coulomb potentials between charged protein residues and the NP surface. Electrostatic potentials were calculated assuming a Debye length of 0.779 nm and Bjerrum length of 0.726 nm as calculated for an electrolyte concentration of 150 mmol in water at 310 K. Binding energies were calculated for radii in the interval 1 – 150 nm and for five values of the electrostatic surface potential between -50 and +50 mV.

## Heat of immersion

A description of how the quantities in Table 5 were calculated has been previously published [31] with the exception of the heats of immersion, which we describe here. The change in enthalpy upon immersion was estimated by conducting simulations of three systems: the surface in contact with water, the surface in a vacuum, and a simulation of bulk water containing an identical number of water molecules as the first system. The immersion enthalpy was calculated from

$$\Delta H_{imm} = \frac{1}{2A}\left(H_{surface-water} - H_{surface-vacuum} - H_{water}\right) \quad (10)$$

where $H$ is the enthalpy of the system and $A$ is the area of the interface. All systems were simulated for 400 ns with 10 ns equilibration.

## Acknowledgements
The work has been funded by EU Horizon2020 under grant agreements No. 686098 (SmartNanoTox project), No. 731032 (NanoCommons project), and 814572 (NanoSolveIT project), and by Science Foundation Ireland through grant 16/IA/4506. The computations of binding free energies were performed on resources provided by the Swedish National Infrastructure for Computing (SNIC) at the Parallel Computer Center (PDC).

**Supporting information**

**Metadynamics calculations**
For computation of adsorption free energies and potentials of mean force at nanomaterials surfaces we used the metadynamics (MetaD) methodology using surface separation distance (SSD) as a collective variable. The SSD was determined as the distance between the sorbent center of mass, and the outermost layer of the surface atoms. The well-tempered MetaD approach[52] was implemented. The binding free energy of a molecule was calculated from a converged adsorption profile $F(s)$ with the estimator

$$\Delta F_{ads} = -k_B T \ln \ln \left( \frac{1}{\delta} \int_0^\delta ds\, e^{-F(s)/k_B T} \right), \qquad (8)$$

where δ is the thickness of the adsorption layer. The result depends weakly (logarithmically) on this parameter so its exact value is not of high importance for the binding free energy. We used $d = 0.8$ nm in our calculations.

All computations have been carried out for systems containing 7000-8000 atoms in a 3-dimensional periodic box of the size about 3×3×8 nm, with 2-dimensional periodic $TiO_2$ slab in *XY* direction and about 1700 water molecules. Gromacs v2018.1[53] with PLUMED plugin v2.5[54] were used in all simulations. The $TiO_2$ slab was prepared by repeating the unit cell the necessary amount of times, and undercoordinated surface Ti atoms were hydrated by either adsorbed water or hydroxyl groups in the ratio 70% to 30%. This fraction (30%) of hydroxyl groups provides surface charge of about −0.65$e$ nm$^{-2}$ which corresponds to the experimentally measured surface charge of $TiO_2$ NPs at neutral pH.[55] The solute molecule was placed outside the material, and the remaining space in the simulation box was filled with water molecules (TIP3P model). The system was first energy-minimized for 10,000 steps using the steepest gradient method. Then the system was equilibrated in the *NVT*-ensemble simulations for 20 ps and in *NPT*-ensemble for 1 ns with time step 1 fs. The temperature was set to 300 K and pressure to 1 bar. Production simulations with metadynamics were carried out in *NVT* conditions, with established in preliminary *NPT* simulations volume, for 300 ns. A *v*-rescale thermostat with relaxation time 1 ps was used to ensure correct ensemble fluctuations. The Particle-mesh Ewald method was used to treat both electrostatic and Lennard-Jones interactions outside the real-space cutoff 10 Å. The motion of the centers of mass of material and solvent were removed separately to avoid artificial flow of the system through periodic boundaries. In order to reduce time spent by the adsorbate in the bulk solvent far from the surface, visiting of such states was prevented by a soft wall potential:

$$U_{wall}(s) = k(s-a)^4 \qquad (9)$$

with the force constant $k = 40$ kJ mol$^{-1}$·Å$^{-4}$ and $a = 1.5$ nm.

**PMFs for TiO$_2$ Anatase 101 and TiO$_2$ Rutile 110**

Here, we present plots of the potentials of mean force (PMFs) calculated for the interaction between the biomolecular fragments and the two TiO$_2$ surfaces.

**Figure S1**: Top: PMFs calculated from atomistic simulations for the various amino acids and lipid fragments against a slab of TiO$_2$ rutile (miller index 110). Bottom: As top, except for the anatase (101) surface.

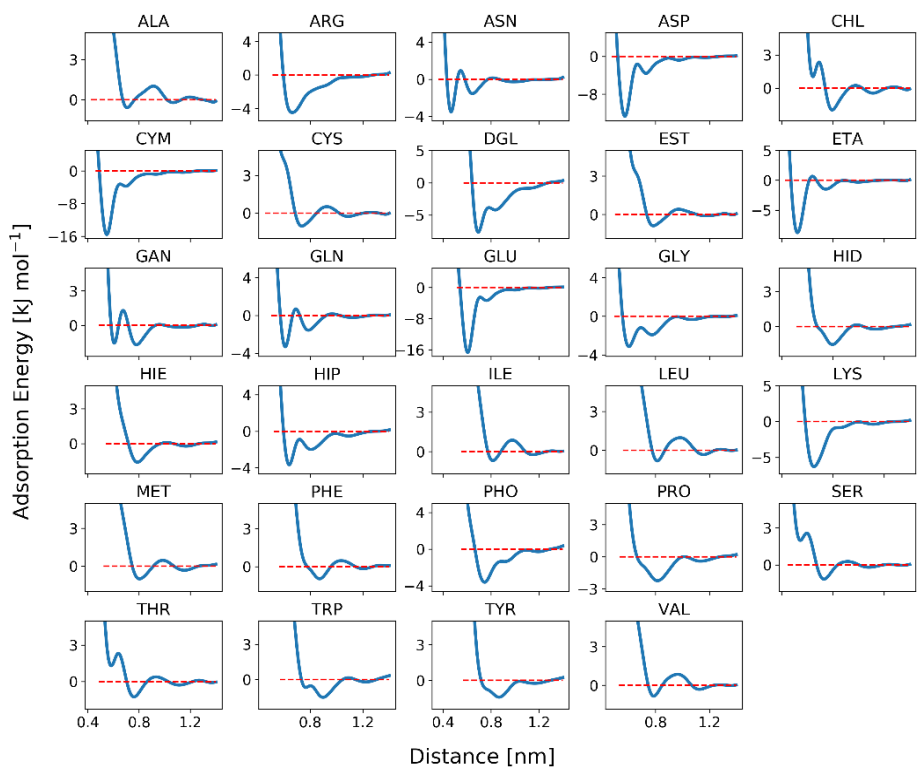

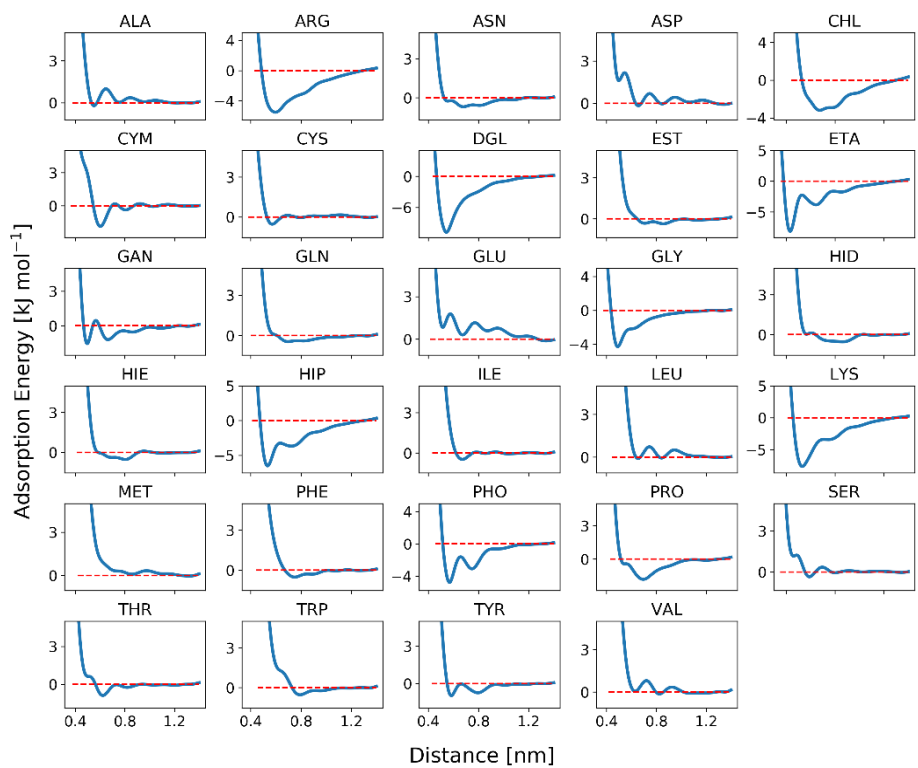

**Protein binding energies**

As Supporting Information, we also provide tabulated data of the calculated descriptors including binding energies for amino acids and a set of 96 proteins on titanium dioxide NPs (anatase(101) and rutile(110)) of a range of zeta potentials (-50- +50 mV) and radii (3 – 150 nm), see file TitaniaNP_Descriptors.xslx.

**References for Supporting Information**